\documentclass[final]{svjour3}
\usepackage{graphicx}
\usepackage{rotating}
\usepackage{amssymb}
\usepackage{mathptmx}
\usepackage{siunitx}
\usepackage{xcolor}
\usepackage[sort,numbers]{natbib}
\makeatletter

\journalname{Journal of Low Temperature Physics}

\bibpunct{[}{]}{,}{n}{}{,}

\begin{document}

\newcommand{\hdblarrow}{H\makebox[0.9ex][l]{$\downdownarrows$}-}
\title{Performance of a low-parasitic frequency-domain multiplexing readout}

\author{A. E. Lowitz$^{1}$ \and A. N. Bender$^{1,2}$ \and P. Barry$^{1,2}$ \and T. W. Cecil$^2$ \and C. L. Chang$^{1,2}$ \and R. Divan$^3$ \and M. A. Dobbs$^4$ \and A. J. Gilbert$^4$ \and S. E. Kuhlmann$^2$ \and M. Lisovenko$^5$ \and J. Montgomery$^4$ \and V. Novosad$^5$ \and S. Padin$^{1,2}$ \and J. E. Pearson$^5$ \and G. Wang$^2$ \and V. Yefremenko$^2$ \and J. Zhang$^2$}

\institute{	$^1$Kavli Institute for Cosmological Physics, University of Chicago, Chicago, IL\\
			$^2$High-Energy Physics Division, Argonne National Laboratory, Argonne, IL\\
			$^3$Center for Nanoscale Materials, Argonne National Laboratory, Argonne, IL \\
			$^4$Department of Physics, McGill University, Montreal, QC \\
			$^5$Materials Sciences Division, Argonne National Laboratory, Argonne, IL\\
}

\maketitle

\begin{abstract}
	
	Frequency-domain multiplexing is a readout technique for transition edge sensor bolometer arrays used on modern CMB experiments, including the SPT-3G receiver. Here, we present design details and performance measurements for a low-parasitic frequency-domain multiplexing readout. Reducing the parasitic impedance of the connections between cryogenic components provides a path to improving both the crosstalk and noise performance of the readout.  Reduced crosstalk will in turn allow higher multiplexing factors. We have demonstrated a factor of two improvement in parasitic resistance compared to SPT-3G hardware.  Reduced parasitics also permits operation of lower-resistance bolometers, which enables better optimization of R$_{\rm{bolo}}$ for improved readout noise performance. The prototype system exhibits noise performance comparable to SPT-3G readout hardware when operating SPT-3G detectors. 


\end{abstract}

\section{Introduction}

	Frequency-domain multiplexing is a technique used by several modern CMB experiments \cite{bender2018,pb2,ebex,litebird, s4_dsr} and other astronomical instruments \cite{sron} to read out many transition edge sensor (TES) bolometers on a single pair of wires.  One inductive-capacitive (LC) filter is connected to each bolometer, assigning that bolometer a unique readout and bias frequency.  Currents flowing through all of the bolometers are first combined and then amplified using a cryogenic amplifier (typically a SQUID array).  The individual signals from each bolometer are later demodulated out of the combined waveform in room temperature digital electronics.  In the digital frequency-domain multiplexing architecture used by SPT-3G (DfMux), 66 bolometers are read out on a single pair of wires \cite{bender2014}.  The carrier signal, after passing through the LC filters and the TES bolometers, is inductively coupled to a series SQUID (superconducting quantum interference device) array (see Fig.~\ref{fig:dfMux}).  To maximize the effective dynamic range of the SQUID, digital active nulling (DAN) is applied at the SQUID input \cite{bender2014, tijmen}.  This `nuller' contains the science signal.  This architecture has demonstrated good on-sky performance with SPT-3G \cite{bender2019, bender2018} with multiplexing factors of up to 68x, but could benefit from improved scalability and electrical performance when applied to future ultra-large focal planes such as those for CMB-S4 \cite{s4_dsr}.

\section{Motivation}
\label{sec:motivation}

\begin{figure} [t]
		\begin{center}
			\begin{tabular}{c} 
				\includegraphics[width=.9\textwidth]{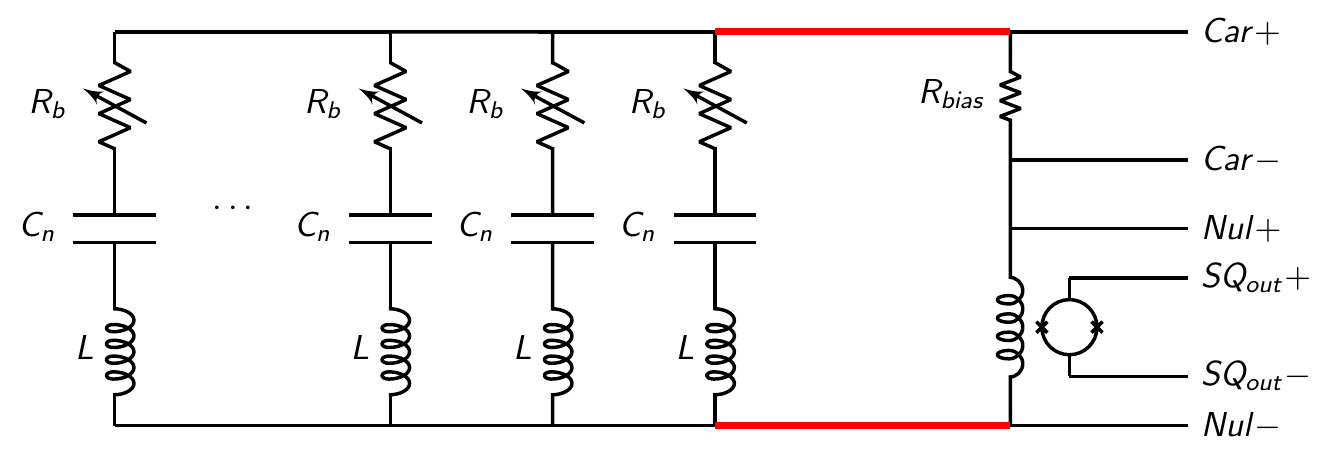}
			\end{tabular}
		\end{center}
		\caption[example] 
		{ \label{fig:dfMux} 
			Schematic of the cryogenic DfMux circuit.  Wiring between the LC chip and the SQUID is indicated by thicker, red lines.  This wiring is a significant source of crosstalk in the cold portion of the DfMux readout, as discussed in Section \ref{sec:motivation}. L and C labels indicate the planar superconducting inductors and capacitors of the LC filters.  The R$_{\rm{b}}$ labels indicate the variable resistance TES bolometers. Labels on the right side indicate the carrier (`Car', i.e. the AC bias), nuller (`Nul'), and SQUID output (`SQ$_{\rm{out}}$') lines.  (Color figure online.) 
		}
	\end{figure}

	In the DfMux system, the LC filters that assign a different bias and readout frequency to each bolometer are on the $\sim$\SI{250}{mK} stage, near the detectors.  A multiplexing unit of bolometers (66 bolometers plus 2 calibration resistors in the case of SPT-3G) is called a module. Wiring between the LC chip and the SQUID consists of broadside-coupled NbTi striplines encapsulated in Kapton and ultrasonically soldered to the LC circuit board on one end and to a breakout circuit board that connects to the SQUID circuit board on the other end \cite{avva}.  This 60 - \SI{75}{cm}-long wiring (shown in thick, red lines in Fig.~\ref{fig:dfMux}), allows the SQUID chip to operate on the \SI{4}{K} stage, physically and thermally distant from the LC filters.  The dominant source of crosstalk in the cryogenic portion of the readout is parasitic impedance in this wiring, which sets up a voltage divider effect such that as the resistance of any bolometer varies according to deposited optical power from the sky, the ratio of the voltage divider is modulated, changing the voltage biases supplied to all other bolometers. This causes changes in TES resistances across the module that mimic those due to optical power deposition \cite{dobbs2012}. Additionally, stable operation of detectors requires the parasitic impedance to be small relative to the TES resistance during operation, therefore the parasitic sourced by the LC-SQUID wiring sets an effective lower limit to the resistance of any TES.  Ability to operate bolometers at lower resistances is desirable because it opens up a larger range of phase space over which R$_{\rm{bolo}}$ can be optimized to improve readout noise performance.  In frequency multiplexing sytems, the optimal bolometer resistance for readout noise performance is typically in the 0.5-\SI{1.0}{\Omega} range, a factor of 2-4 lower than SPT-3G bolometers.  

	The primary goal of this work is to develop a modified DfMux architecture which minimizes the parasitic impedance of the wiring between the LC filters and the SQUID. This will substantially reduce crosstalk, as well as take a first step towards the operation of lower-resistance bolometers and improved readout noise performance.

\section{Design}

\begin{figure} [t]
		\begin{center}
			\begin{tabular}{c} 
				\includegraphics[height=3.5cm]{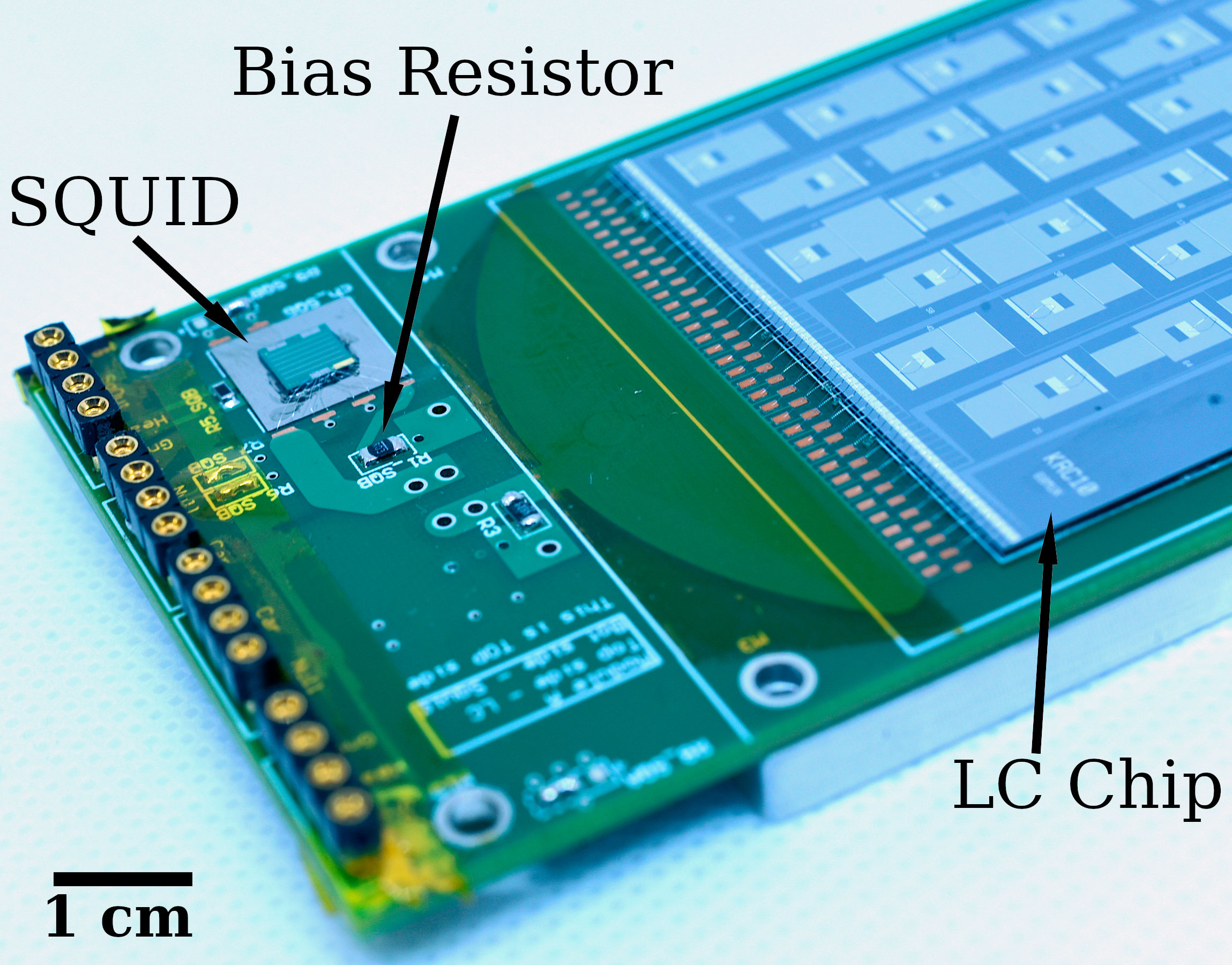}  \includegraphics[height=3.5cm]{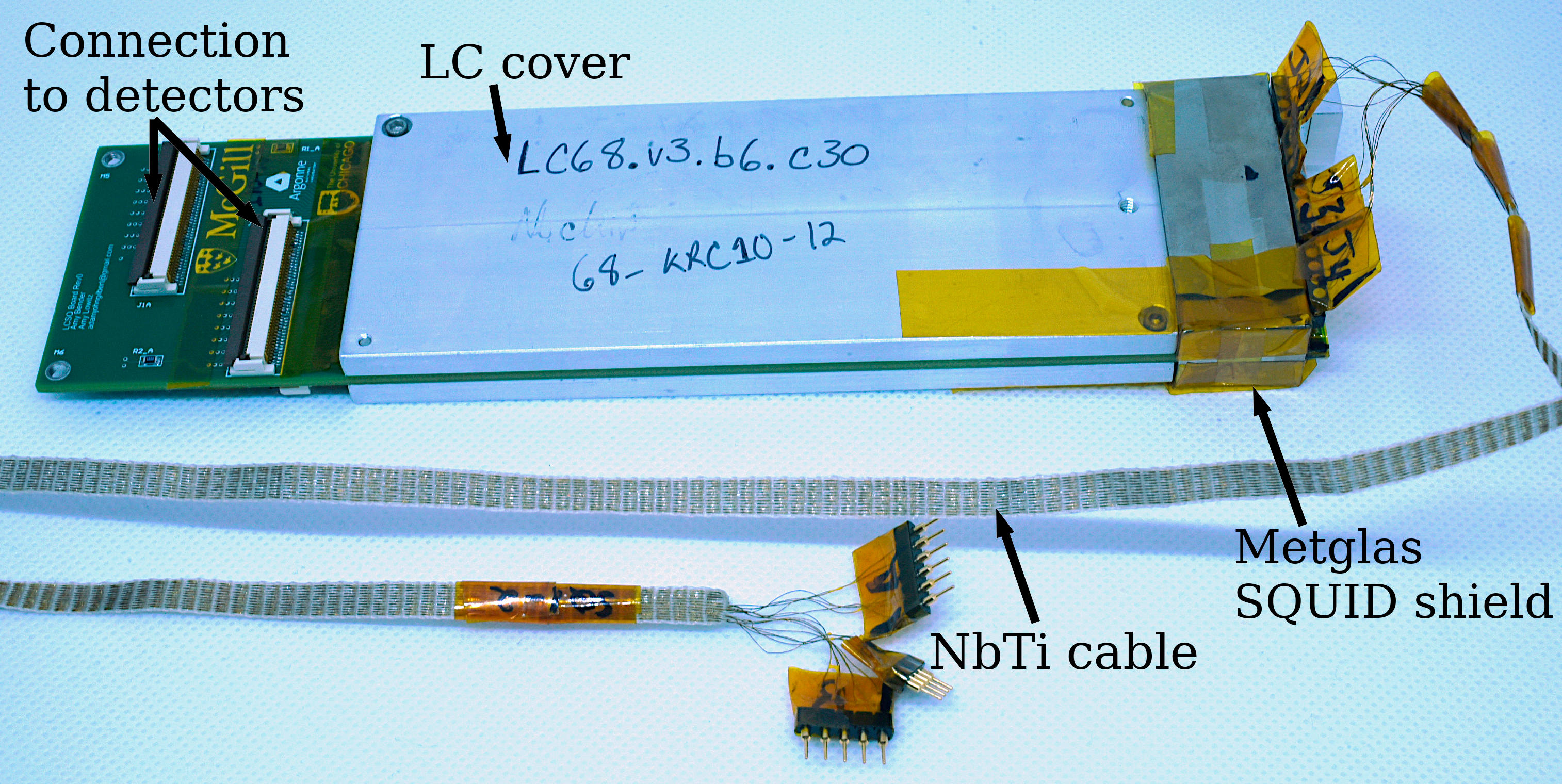}
			\end{tabular}
		\end{center}
		\caption[example] 
		{ \label{fig:photos} 
		{\it Left:} Partially-assembled prototype circuit board showing the SQUID chip, LC chip, bias resistor, and SIP connectors. {\it Right:} Fully-assembled prototype circuit board showing the Metglas cover for the SQUIDs, aluminum cover for the LC chips, and NbTi twisted pair ribbon cable which connects the board to a wiring feedthrough at \SI{4}{K}. (Color figure online.)
		}
	\end{figure}

	The wiring between the SQUID and the LC chip should be as short and wide as possible to reduce the parasitic impedance.  One straightforward way to achieve this is to move the SQUID from the \SI{4}{K} stage to the \SI{250}{mK} stage, onto the same circuit board as the LC chip and to use wide wiring traces. Refer to Fig.~\ref{fig:photos} for photographs of the prototype circuit board design. The design of the prototype hardware retains as much of the existing DfMux design as possible.  This reaps the benefits of reduced parasitic impedance, reduced crosstalk, and improved scalability, while inheriting a relatively high level of technological maturity from the existing technology.  To that end, the prototype circuit board is identical to the SPT-3G LC board from the input to the LC chip through to the connections to the detector wafer.  A small extension was added to the input end of the board to accommodate the SQUID chip and its circuitry.  This design minimizes the wiring length between the SQUID and LC chips, significantly reducing parasitics from that section of wiring.  A single circuit board houses one LC chip and one SQUID chip on each side.  Additionally, the extension to the board is short enough that the prototype hardware fits in every SPT-3G cryogenic testbed (and in principle, in the telescope itself) as a drop-in replacement for the standard SPT-3G DfMux readout hardware. This enables rapid and straightforward testing of the prototype architecture alongside SPT-3G hardware for comparison.

	SQUIDs are sensitive magnetometers and require careful magnetic shielding in order to operate optimally as an amplifier \cite{squidHandbook}.  An aluminum box (for mechanical protection) covered with six layers of Metglas \cite{metglas} (a high-permeability material used for magnetic shielding) covers the SQUIDs (see Fig.~\ref{fig:photos}).  Comparison testing has shown that the Metglas box performs as well as an Ammuneal box for magnetic shielding in this application~\cite{lowitzSPIE}.  A separate aluminum cover acts as a mechanical shield for the LC chips.

	A convenient additional benefit of removing the long wiring between the LC chip and the SQUID chip is that it eliminates a complicated, expensive, and labor intensive component of the cryogenic wiring.  Long wiring is still required to reach from the detector stage to the \SI{4}{K} stage.  However, since the long wiring in the prototype system is between the SQUID output and the warm electronics (instead of between the SQUID and the LC), the impedance of this long wiring does not contribute to crosstalk.  Therefore, the technical requirements for this wiring become significantly less restricted.  The current prototype uses commercially-available NbTi twisted-pair woven loom ribbon cable, hand-connectorized with SIP (single inline package) socket connectors.  Future revisions may use the same NbTi ribbon cable with commercially-potted connectors for improved scalability.

\section{Performance}

	\subsection{SQUID Performance}
	
\begin{figure} [ht]
			\begin{center}
				\begin{tabular}{c} 
					\includegraphics[height=4.13cm]{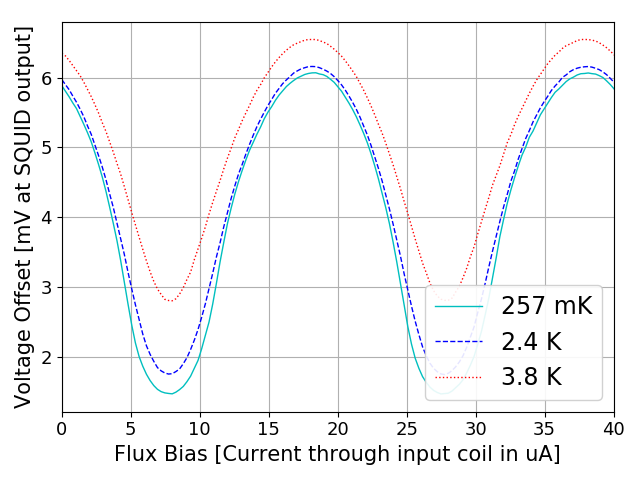}  
					\includegraphics[height=4.13cm]{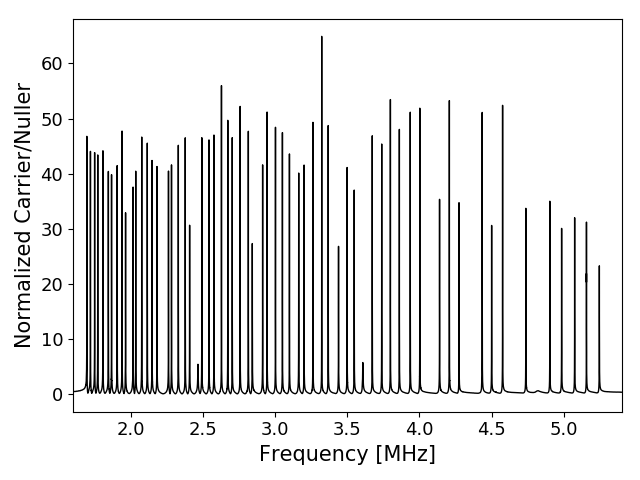}\\ 
					\includegraphics[height=4.13cm]{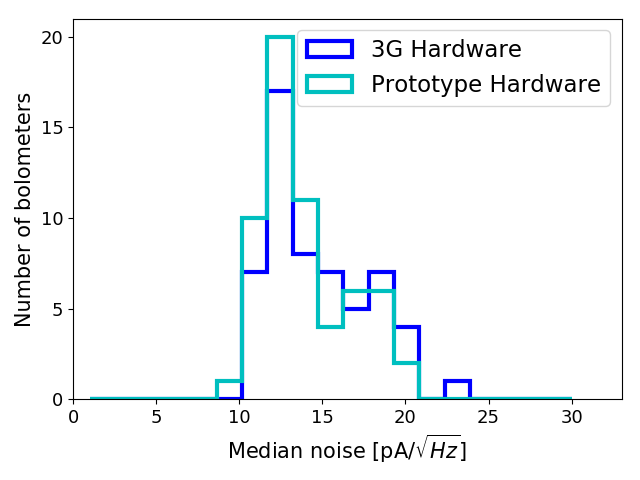}  
					\includegraphics[height=4.13cm]{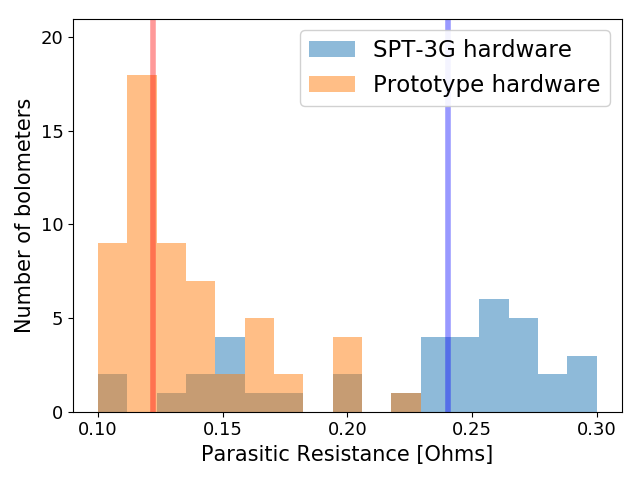}
				\end{tabular}
			\end{center}
			\caption[example] 
			{ \label{fig:performance}
				\textbf{\textit{Top Left:}} Performance of the NIST SA13 SQUIDs is similar across a range of temperatures.  Shown here is the flux bias vs the SQUID voltage offset for the same SQUID chip at its optimal (maximal peak-to-peak voltage) current bias. For this SQUID chip, the transimpedances were Z=\SI{945}{\Omega}, Z=\SI{850}{\Omega}, and Z=\SI{683}{\Omega} at \SI{257}{mK}, \SI{2.4}{K}, and \SI{3.8}{K} respectively.  \textbf{\textit{Top Right:}} Network analysis (carrier signal divided by nuller signal, across the full readout frequency range) of the prototype module showing 63 resonance peaks. \textbf{\textit{Bottom Left:}} A comparison of noise performance for the prototype system and the SPT-3G readout hardware with the detectors at T $>$ T$_{\rm{c}}$. The prototype system clearly demonstrates comparable performance. \textbf{\textit{Bottom Right:}} Shown here is the parasitic resistance of one module of bolometers on prototype hardware and one module on SPT-3G hardware.  The orange and blue vertical lines indicate the medians for the prototype and SPT-3G hardware respectively. The reduced wiring length between the SQUID chip and LC chip substantially reduces the parasitic resistance in the prototype system compared to standard DfMux readout hardware.  (Color figure online.)
			}
		\end{figure}

		The prototype system has been tested using NIST SA13 SQUID arrays.  Their performance at sub-Kelvin temperatures is comparable to or slightly better than their performance at more typical operating temperatures of 2 - \SI{4}{K}.  Fig.~\ref{fig:performance} ({\it top left}) shows a comparison plot of the flux bias versus output voltage for a representative SA13 at three temperatures.  Performance is very similar at all three temperatures, with a marginal increase in peak-to-peak voltage and transimpedance at lower temperatures.  
		
		Thermal dissipation is also an important consideration when putting SQUIDs on the same thermal stage as detectors. The thermal dissipation of a SQUID can be estimated as the product of the current bias and the voltage offset at the SQUID output. Tuned at a typical operating current- and flux- bias, a single representative SA13 (able to operate one module of bolometers) dissipates about \SI{150}{nW} of heat onto the detector stage.  At this level of thermal dissipation, operating small numbers of SA13s works well, but scaling up to the hundreds of SQUIDs required for operation of a large detector array is not feasible using a typical sorption refrigerator.  For example, a Chase Cryogenics \cite{chase} gaslight sorption refrigerator provides $\sim$\SI{4}{\mu}W of cooling power on the $\sim$\SI{250}{mK} stage.  For scalability, either a dilution refrigerator or a SQUID with lower thermal dissipation will be required.

	\subsection{Bolometer Operation and Noise Performance}
	
		The detector wafer used in all of the tests described in this work is an SPT-3G wafer that was successfully observed with in 2017.  When comparing the prototype to SPT-3G hardware, the SPT-3G hardware data is laboratory test data taken on the same detector wafer at the same time or nearly the same time as the prototype hardware data.  A side-by-side measurement of the prototype and SPT-3G DfMux hardware parasitic resistance demonstrates a factor of two improvement in median parasitic resistance, as shown in Fig.~\ref{fig:performance} ({\it bottom right}).  This demonstration of decreased parasitic resistance paves the way to operating lower-resistance bolometers.  
		
\begin{figure} [t]
			\begin{center}
				\begin{tabular}{c} 
					\includegraphics[width=.9\textwidth]{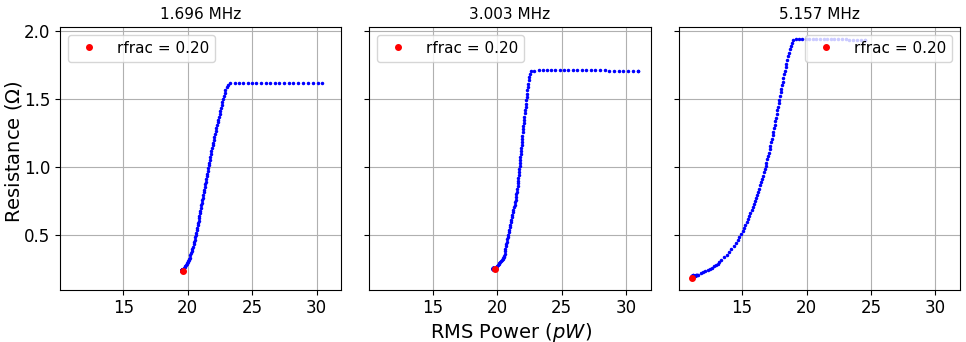}
				\end{tabular}
			\end{center}
			\caption[example] 
			{ \label{fig:RP} 
				Operation of a full module of 57 bolometers down to a fractional resistance of $\rm{R_{frac}}=0.2\times \rm{R_{n}}$ has been demonstrated with the prototype.  Shown are resistance vs power curves for three representative bolometers as they drop into the superconducting transition.  Each bolometer tuning step from maximum bias power down into the superconducting transition is indicated by a small blue dot and the final tuned point is indicated by a large red dot.  These curves have been corrected to remove the parasitic resistance.  (Color figure online.) 
			}
		\end{figure}

		Of a possible 66 bolometers connected to this module, 58 should be operable in our current testing configuration.  The eight missing bolometers are known flaws on the detector wafer and LC chip, identified by looking for shorts, opens, and missing peaks in the network analysis.  A network analysis (carrier signal divided by nuller signal; a tool for identifying working channels and measuring their frequency) showing the LC resonant peaks is shown in Fig.~\ref{fig:performance}. Of the full module of 58 bolometers that can be biased, we have successfully placed 57 into the TES superconducting transition to a fractional resistance of $0.2\times \rm{R_{n}}$ and held them there with good stability under dark, steady-state conditions. Fig.~\ref{fig:RP} shows resistance versus power curves for a representative selection of bolometers as they are dropped into the TES superconducting transition. For comparison, under the same conditions, the standard SPT-3G readout hardware was able to operate at no lower than $0.3\times \rm{R_{n}}$ before the bolometers became too unstable and the whole module latched superconducting. The improved stability at very low fractional resistances likely results from the decreased parasitic resistance in the prototype system, but more investigation is required to develop a detailed understanding of the performance improvements.
		
		The median noise of the system is measured with the detectors at T $>$ T$_{\rm{c}}$, providing an effective diagnostic of the level of readout noise.  Fig.~\ref{fig:performance} ({\it bottom left}) shows that the noise in the prototype system is comparable to that of the SPT-3G hardware. This is also comparable to the full, fielded readout noise performance of the SPT-3G array~\cite{bender2019}. 
		
		While there is not yet a detailed measurement of the parasitic inductance in the prototype system, an estimate based on the geometry of the traces between the SQUID and the LC on the prototype circuit board is 5-\SI{10}{nH}. From this parasitic inductance value, the effect on crosstalk can be simulated.  For an SPT-3G-like system, which has measured parasitic inductance of \SI{45}{nH} \cite{avva}, the simulated median nearest-neighbor crosstalk is 0.58\%.  Reducing the parasitic inductance to \SI{10}{nH}, a conservative estimate for the prototype hardware, reduces the simulated median nearest-neighbor crosstalk to 0.17\%.  
		

\section{Conclusions and Future Work}

We have demonstrated operation of a full module of frequency-multiplexed bolometers using a prototype system with parasitic impedance between the LC and SQUID significantly reduced compared to existing systems. Readout noise performance is comparable to that of SPT-3G hardware.  NIST SA13 SQUIDs perform well overall at sub-Kelvin temperatures. However, to improve scalability by enabling operation of a large number of SQUIDs on the cold stage, either dilution refrigeration or SQUIDs with lower thermal dissipation will be required.  Finally, we have demonstrated a reduction in parasitic resistance of a factor of two compared to SPT-3G readout hardware.  This improvement in R$_{\rm{parasitic}}$ enables operation of bolometers with lower resistance than SPT-3G, which opens up a larger range of phase space for optimizing R$_{\rm{bolo}}$ based on noise performance.  Broadly, the prototype architecture shows promise as a more scalable, potentially higher-multiplexing-factor, alternative to the standard DfMux system, and more detailed characterization and laboratory testing is warranted to fully mature this technology.

Ongoing testing of this prototype hardware will focus on detailed characterization of the in-transition noise level and crosstalk between bolometers.  Additionally, we plan to make a detailed measurement of the thermal dissipation of the SA13 SQUIDs in this configuration and will identify a design requirement for SQUID thermal dissipation.  We also plan to test the prototype architecture with lower resistance bolometers and scale up to operating a larger number of modules.

\begin{acknowledgements}
	
	Work at the University of Chicago is supported by the National Science Foundation through grant PLR-1248097.  Work at Argonne National Lab is supported by UChicago Argonne LLC, Operator of Argonne National Laboratory (Argonne). Argonne, a U.S. Department of Energy Office of Science Laboratory, is operated under contract no. DE-AC02-06CH11357.  The McGill authors acknowledge funding from the Natural Sciences and Engineering Research Council of Canada, Canadian Institute for Advanced Research, and the Fonds de recherche du Québec Nature et technologies.

\end{acknowledgements}

\pagebreak

\end{document}